\documentclass[12pt]{article}

\begin{document}

\begin{center}
\textbf{A\ POTENTIAL\ OF\ INTERACTION\ BETWEEN\ TWO- AND\ THREE-DIMENSIONAL\
SOLITONS}

Boris A. Malomed\footnote{%
electronic address malomed@eng.tau.ac.il}

\vspace{1.0cm}Department of Interdisciplinary Studies, Faculty of
Engineering, Tel Aviv University, Tel Aviv 69978, Israel
\end{center}

\newpage

\begin{center}
\textbf{ABSTRACT}
\end{center}

A general method to find an effective potential of interaction between far
separated 2D and 3D solitons is elaborated, including the case of 2D vortex
solitons. The method is based on explicit calculation of the overlapping
term in the full Hamiltonian of the system (\emph{without} assuming that the
``tail'' of each soliton is not affected by its interaction with the other
soliton). The result is obtained in an explicit form that does not contain
an artificially introduced radius of the overlapping region. The potential
applies to spatial and spatiotemporal solitons in nonlinear optics, where it
helps to solve various dynamical problems: collisions, formation of bound
states (BS's), etc. In particular, an orbiting BS of two solitons is always
unstable. In the presence of weak dissipation and gain, the effective
potential can also be derived, giving rise to bound states similar to those
recently studied in 1D models.

PACS numbers: 42.50.Rh; 42.50.Ne; 42.65.Vh; 52.35.Sb; 03.40.Kf \newpage

\section{INTRODUCTION}

Recent progress in studies of two-dimensional (2D) solitons in models of
non-Kerr nonlinear optical media has attracted a lot of interest to their
interactions. 2D vortex solitons and interactions between them in the
quintic nonlinear Schr\"{o}dinger (QNLS) equation were studied in \cite
{Manolo}, nonplanar interactions between 2D solitons in a medium with the
quadratic ($\chi ^{(2)}$) nonlinearity were considered, numerically and
analytically, in \cite{SKB}, and various features of the interaction between
2D solitons in photorefractive media were revealed by numerical simulations
and direct experiments \cite{Moti,Moti2,Krolik}. The nonlinearity must be
non-Kerr because the usual cubic (Kerr) self-focusing term gives rise to
collapse in 2D and 3D (three-dimensional) cases. As it was demonstrated in 
\cite{KR}, the collapse does not take place in any physical dimension in the
model with the $\chi ^{(2)}$ nonlinearity. This opens way to stable 2D and
3D \emph{spatiotemporal} solitons, or ``light bullets'' (LBs) \cite{Yaron}.
The $\chi ^{(2)}$ LBs were recently studied in detail in \cite{Aussie} and 
\cite{Torner}.

The objective of this work is to find an effective potential of interaction
between 2D and 3D solitons in isotropic media (note that, as it was
demonstrated in a very recent experimental work \cite{Moti2}, the
interaction of 2D solitons in intrinsically anisotropic photorefractive
media is, in effect, practically isotropic too). The interaction potential
is necessary to solve various dynamical problems, such as collisions,
formation of bound states of solitons, etc., including a practically
important problem of designing all-optical switching by means of interaction
between 2D optical solitons. It will be demonstrated that a universal\emph{\ 
}effective potential can be obtained analytically by means of a technique
which generalizes that developed for the 1D solitons in \cite{BS}. As a
paradigm model, one can take the multidimensional \emph{quintic
Ginzburg-Landau} (QGL) equation, 
\begin{equation}
iv_t+\frac 12\nabla ^2v+|v|^2v-\alpha |v|^4v=-iv+i\gamma _1\nabla
^2v+i\gamma _2|v|^2v-i\gamma _3|v|^4v,  \label{GL}
\end{equation}
where the coefficients $\alpha $ and $\gamma _{1,2,3}$ are positive. The
QNLS equation is a conservative version of (\ref{GL}), without its
right-hand side. The quintic defocusing term $\sim \alpha $ is included in
order to prevent the collapse. Note that this term is not merely the
simplest one that stabilizes the model: according to experimental data \cite
{PTS}, the combination of the focusing cubic and defocusing quintic terms
adequately models the nonlinear optical properties of some real materials.
The first two terms on the right-hand side of (\ref{GL}) take into regard
linear losses, the cubic term $\gamma _2$ accounts for \emph{nonlinear gain}
which compensates the losses, and the quintic dissipation term $\sim \gamma
_3$ provides for the overall stabilization of the model. The QGL equation
was first introduced in \cite{PS} (in the 2D form), and its 1D
(one-dimensional) version has later attracted a great deal of interest (see,
e.g., \cite{CH} and references therein). In particular, stable localized
pulses in the 1D QGL equation were found in \cite{1D} for the case of weak
dissipation (relevant for the applications to nonlinear optics), $0\leq
\gamma _{1,2,3}\ll 1$, that will be also assumed here. The existence of the
stable pulses in the opposite limit of strong dissipation was independently
shown in three different works \cite{3}. Actually, the model (\ref{GL}) is
selected just for the reference, as the one that certainly gives rise to
stable multidimensional solitons; as it will be seen below, the derivation
of the effective potential for the interaction between the solitons,
presented in this work, is quite universal and may be applied to any
conservative or weakly dissipative model that supports multidimensional
solitons.

Note that stable 2D solitons, as well as two-soliton bound states, were also
found numerically in a model with the quintic nonlinearity similar to that
in Eq. (\ref{GL}), in which, however, the linear part is of a higher order,
containing the operators $\partial ^2/\partial t^2$ and $\nabla ^4$ \cite
{Aranson}. However, that model is essentially more complicated than (\ref{GL}%
), and its physical applications are less clear.

The paper is organized as follows. In section 2, the 2D and 3D soliton
solutions are briefly considered, with emphasis on the form of their
asymptotic ``tails'', which determine the effective interaction potential.
In the same section, the model (\ref{GL}) is also reformulated in terms of
nonlinear optics, where it finds applications of two types: the description
of spatial cylindrical solitons in the bulk medium, and 2D and 3D LBs in
the, respectively, 2D nonlinear waveguide or 3D bulk. The multidimensional
solitons in the model with the $\chi ^{(2)}$ nonlinearity have their own
peculiarities, which are summarized in a separate subsection in section 2.
In section 3, the interaction potential is analytically derived, in a
general form, for the 2D and 3D solitons. In the same section, the
interaction potential for LBs is also considered. In particular, the
potential may be \emph{spatiotemporally anisotropic} for the $\chi ^{(2)}$
LBs, while in the other models it can always be cast into an effectively
isotropic form. Concluding remarks are collected in section 4, including a
discussion of a possibility of existence of bound states of the solitons. In
particular, it is concluded, in accord with the recent results obtained for
the $\chi ^{(2)}$ spatial solitons in \cite{SKB}, that a bound state of two
solitons orbiting around each other may exist (in the dissipationless
model), but it is always unstable. In the presence of the weak dissipation
and gain, there are bound states of quiescent solitons, quite similar to
those recently studied in the 1D model (that may be both unstable and almost
stable). New possible states in the 2D and 3D cases are soliton lattices and
``molecules''.

\section{TWO- AND\ THREE-DIMENSIONAL\ SOLITONS}

\subsection{The General Case}

A general stationary solution to Eq. (\ref{GL}) is $v=\exp \left( -i\omega
t\right) V(\mathbf{r})$, where $V(\mathbf{r})$ satisfies the equation 
\begin{equation}
\frac 12\nabla ^2V+|V|^2V-\alpha |V|^4V+\omega V=-iV+i\gamma _1\nabla
^2V+i\gamma _2|V|^2V-i\gamma _3|V|^4V.  \label{stationary}
\end{equation}
In the 2D case, the solution is restricted to the form 
\begin{equation}
V(\mathbf{r})=\exp \left( is\theta \right) \;\mathcal{V}(r),\;s=0,\pm 1,\pm
2,...\;,  \label{sol}
\end{equation}
where $r$ and $\theta $ are the polar coordinates, $s\neq 0$ corresponding
to a \emph{vortex soliton}, and $\mathcal{V}(r)$ exponentially decays at $%
r\rightarrow \infty $. From the consideration of Eq. (\ref{GL}) it follows
that the asymptotic form of the soliton at $r\rightarrow \infty $ is 
\begin{eqnarray}
\mathcal{V}(r) &\approx &A_sr^{-1/2}\exp \left( -\kappa r\right) ,
\label{2D} \\
\kappa &=&\sqrt{-\frac{\omega +i}{1/2+i\gamma _1}}\approx \sqrt{-2\omega }%
-iq,\;q=\frac 1{\sqrt{-2\omega }}+\gamma _1\sqrt{-2\omega },  \label{kappa}
\end{eqnarray}
and, at $r\rightarrow 0$, 
\begin{equation}
\mathcal{V}(r)\approx a_sr^{|s|}  \label{hole}
\end{equation}
(i.e., the vortex soliton has a hole in its center), with unknown constants $%
A_s$ and $a_s$. The expansion of $\kappa $ in (\ref{kappa}) employs the fact
that, in the weakly dissipative regime, $\gamma _1$ is small, and $\omega
\gg 1$, as the dissipation coefficient in front of the term $-v$ in Eq. (\ref
{GL}) is $1$.

The stability of the $s=0$ soliton in the model (\ref{GL}) is very
plausible, and, in the conservative version of (\ref{GL}), the stability of
the vortex soliton with $|s|=1$ was numerically demonstrated in \cite{Manolo}%
. It is not known if the solitons with $|s|>1$ are stable (note that all the
bright vortex solitons are unstable in the $\chi ^{(2)}$ model, see, e.g., 
\cite{Ll}). Below, an arbitrary integer value of $s$ will be kept, as the
potential can be derived in the general case, provided that the two solitons
have $s_1=\pm s_2$.

Description of 3D solitons with the internal ``spin'' is a rather
complicated problem, therefore only the 3D solitons with the zero spin will
be considered here. The corresponding solution is sought for in the form of
Eq. (\ref{sol}) with $s=0$, and with the difference that $r$ is now the
radial variable in the 3D space, hence 
\begin{equation}
\mathcal{V}(r)\approx A_sr^{-1}\exp \left( -\kappa r\right)  \label{3D}
\end{equation}
at $r\rightarrow \infty $.

In the conservative version of the model, the frequency $\omega <0$ is an
arbitrary parameter of the soliton, while the amplitude $A_s$, that can be
found numerically, is a function of $\omega $ (as well as $a_s$ in Eq. (\ref
{hole})). In the presence of the weak dissipation and gain, an actual
soliton solution is selected from the continuous family as the one providing
for a balance of the ``number of photons'', $\int_0^\infty |V(r)|^2r^{D-1}dr$
\cite{1D}. In that case, the value of $\omega $ should also be found
numerically. Below, $\omega $ and $A_s$ will be treated as given parameters.

In the application to the nonlinear optics, the QNLS version of the 2D model
(\ref{GL}) describes time-independent light distributions in a 3D medium, so
that the variable $t$ is not time, but the propagation coordinate. The
dynamics of LBs in 2D and 3D optical media is governed by an equation that
is also similar to Eq. (\ref{GL}). Neglecting the dissipative part, the
corresponding QNLS equation is 
\begin{equation}
iv_z+\frac 12\left( \nabla _{\bot }^2v+v_{\tau \tau }\right) +|v|^2v-\alpha
|v|^4v=0,  \label{anomalous}
\end{equation}
where $v$ is the envelope of the electromagnetic waves, $z$ and $\tau \equiv
t-z/c_{\mathrm{gr}}$ are the propagation coordinate and the so-called
retarded time, $c_{\mathrm{gr}}$ being the mean group velocity of the
carrier wave, and the operator $\nabla _{\bot }^2$ acts on the transverse
coordinate(s). In Eq. (\ref{anomalous}), \emph{anomalous} temporal
dispersion (accounted for by the term $v_{\tau \tau }$) is assumed. A
spatiotemporal-soliton solution to Eq. (\ref{anomalous}) (i.e., LB) can be
sought for in the form (cf. Eq. (\ref{sol})) 
\begin{equation}
v=\exp \left( ikz\right) \;\mathcal{V}(\xi ),\;\xi \equiv \sqrt{r_{\bot
}^2+\tau ^2}.  \label{LB}
\end{equation}
Here, $k$ is the \textit{propagation constant} and $r_{\bot }$ is the
transverse coordinate in the 2D model, or the radial variable in the
transverse plane in the 3D model. In the latter case, a more general
solution with a ``spatiotemporal spin'' can be looked for in the form 
\begin{equation}
v=\exp \left( ikz+is\theta \right) \;\mathcal{V}(\xi ),\;  \label{LBhole}
\end{equation}
where this time $\theta $ is the formal angular coordinate on the plane ($%
r_{\bot }$,$\tau $). The solution (\ref{LBhole}) has a ``hole'' in its
center, cf. Eq. (\ref{hole}). The asymptotic form of all the LB solutions at 
$\xi \rightarrow \infty $ is similar to that given above by Eqs. (\ref{2D}),
(\ref{kappa}) and (\ref{3D}): 
\begin{equation}
\mathcal{V}(\xi )\approx A\xi ^{-(D-1)/2}\exp \left( -\sqrt{2q}\xi \right) .
\label{2D,3D}
\end{equation}

\subsection{The model with the quadratic nonlinearity}

An allied physically important model is that describing multidimensional $%
\chi ^{(2)}$ media \cite{Aussie}: 
\begin{eqnarray}
iv_z+\frac 12\left( \nabla _{\bot }^2v+v_{\tau \tau }\right) -v+v^{*}w &=&0,
\label{FH} \\
2iw_z+\frac 12\left( \nabla _{\bot }^2w+\delta w_{\tau \tau }\right) -\gamma
w+\frac 12v^2 &=&0,  \label{SH}
\end{eqnarray}
where $v$ and $w$ are envelopes of the fields at the fundamental harmonic
(FH) and second harmonic (SH), $\gamma >0$ is a mismatch parameter, and $%
\delta $ is a \emph{relative coefficient} of the temporal dispersion. In the
real physical situations, $\delta <1$, including negative values (which
correspond to the normal dispersion at the second harmonic). As it was shown
in \cite{Aussie} and \cite{Torner}, the spatiotemporal soliton solutions to
Eqs. (\ref{FH}) and (\ref{SH}) can only exist if $\delta \geq 0$. However,
the solution cannot be sought for in the form (\ref{LB}), except for the
unrealistic special case $\delta =1$. The asymptotic form of the soliton can
be nevertheless easily found from the linearized versions of Eqs. (\ref{FH})
and (\ref{SH}), cf. Eqs. (\ref{2D})--(\ref{3D}) and (\ref{LB}): 
\begin{eqnarray}
v_{\mathrm{s}} &\approx &A\xi ^{-(D-1)/2}\exp \left( -\sqrt{2}\xi \right) ,
\label{vasymp} \\
w_{\mathrm{s}} &\approx &B\widetilde{\xi }^{-(D-1)/2}\exp \left( -\sqrt{%
2\gamma }\widetilde{\xi }\right) ,\;\widetilde{\xi }^2\equiv r_{\bot
}^2+\delta ^{-1}\tau ^2.  \label{wasymp}
\end{eqnarray}
Here, only the case $s=0$ is considered, and the propagation constant is not
explicitly introduced, as it may be absorbed by the mismatch parameter $%
\gamma $.

The consideration of Eq. (\ref{SH}) readily demonstrates that, while the
asymptotic expression (\ref{vasymp}) for FH is always relevant, the
expression (\ref{wasymp}) makes sense only if it decays at $r_{\bot },\tau
\rightarrow \infty $ \emph{not faster} than $v_{\mathrm{s}}^2$. Further
straightforward analysis shows that this condition is always satisfied,
provided that $\gamma <4\delta $, and never satisfied, if $\gamma >4$. In
the intermediate case 
\begin{equation}
4\delta <\gamma <4  \label{inter2}
\end{equation}
(recall that the physical constraint is $\delta <1$, and $\gamma =4$ has a
special meaning corresponding to the \textit{exact matching} between FH and
SH \cite{Aussie}), the condition holds, on the plane of the variables $r$
and $\tau $, inside the sector 
\begin{equation}
(\tau /r_{\perp })^2<\frac{4-\gamma }{\gamma -4\delta }\delta ,
\label{sector2}
\end{equation}
and does not hold outside this sector. The $\chi ^{(2)}$ solitons for which
this condition holds, i.e., the shape of their asymptotic ``tails'' in each
harmonic is \emph{independently} determined by the corresponding linearized
equations, may be naturally called \emph{free-tail solitons}.

In the case when the above condition does not hold, i.e., $v_{\mathrm{s}}^2$%
, as given by Eq. (\ref{vasymp}), decays slower than $w_{\mathrm{s}}$ as per
Eq. (\ref{wasymp}), the latter expression does not apply. In this case, the
actual decay of the SH tail is governed by the quadratic term in Eq. (\ref
{SH}), while Eq. (\ref{vasymp}) remains valid. The final result which, in
this case, replaces Eq. (\ref{wasymp}) is 
\begin{equation}
w_{\mathrm{s}}\approx \frac 12A^2\frac{\xi ^{3-D}}{(\gamma -4)\xi
^2+4(1-\delta )\tau ^2}\exp \left( -2\sqrt{2\gamma }\xi \right)
\label{locked}
\end{equation}
(note that the underlying conditions that define the present case guarantee
that the expression (\ref{locked}) is not singular). The solitons of this
type may be called, on the contrary to the free-tail ones defined above, 
\emph{tail-locked} solitons. Note that, in the intermediate case (\ref
{inter2}), the free-tail asymptotic expression (\ref{wasymp}) holds inside
the sector (\ref{sector2}), while the tail-locked expression (\ref{locked})
is valid outside the sector.

\section{THE INTERACTION\ POTENTIAL}

\subsection{The General Case}

Coming back to the paradigm model (\ref{GL}), one notes that the
conservative (left-hand) side of its stationary version (\ref{stationary})
can be derived from the Hamiltonian 
\begin{equation}
H=\int \left( \frac 12\left| \nabla V\right| ^2-\frac 12|V|^4+\frac 13\alpha
|V|^6-\omega |V|^2\right) d\mathbf{r}.  \label{H}
\end{equation}
The Hamiltonian allows one to define an effective interaction potential for
two separated solitons \cite{GO}. In the original works, the wave field
corresponding to the two-soliton configuration was postulated to be a linear
superposition of two isolated solitons. This was substituted into the
Hamiltonian, and a term produced by the overlapping of the ``body'' of each
soliton with a weak ``tail'' of the other one was identified as an effective
interaction potential. This approach requires actual calculation of the
corresponding integral term in (\ref{H}), a drawback being that a distortion
of the ``tail'' due to its interaction with the other soliton is ignored. In
this work, a more consistent approach will be developed, following that
elaborated for the 1D solitons in \cite{BS}. The method is based on
representing the wave field in a vicinity of each soliton in the form 
\begin{equation}
v(\mathbf{r,}t)=\exp \left( -i\omega t\right) \left[ V_{\mathrm{s}}(\mathbf{r%
})+V_{\mathrm{t}}(\mathbf{r})\right] ,  \label{overlap}
\end{equation}
where $V_{\mathrm{s}}(\mathbf{r})$ is the isolated soliton (\ref{sol}), $V_{%
\mathrm{t}}(\mathbf{r})$ is a small tail generated by the second soliton,
and the influence of a given soliton on the other soliton's ``tail'' is 
\emph{not} neglected. The distance $R$ between the centers of the two
solitons is assumed to be essentially larger than the soliton's size $\sim
\kappa ^{-1}$, see Eq. (\ref{2D}). A similar structure of the wave field is
assumed near the center of the second soliton.

Only the case when the interacting solitons are identical is considered (in
particular, $s_1=\pm s_2$ for the 2D vortex solitons, and the amplitudes $%
|A_s|$, defined as per Eq. (\ref{2D}), are equal), hence the solitons have
the same frequency $\omega $, which allows one to define a phase difference $%
\psi $ between them. The case of the identical solitons is the most relevant
one, as parameters of the solitons are, in a real physical situation,
uniqely selected by the above-mentioned balance between the gain and
dissipation.

The next step is, as it was said above, to insert the expression (\ref
{overlap}) into the Hamiltonian (\ref{H}) and calculate the overlap term in
an area around the first soliton, adding then a symmetric contribution from
the vicinity of the other soliton. In the first approximation, only the
terms linear in $V_{\mathrm{t}}$ are to be kept, which yields the following
expression for the effective interaction potential, 
\begin{eqnarray}
U_D(R,\psi ) &=&[\int \left( \frac 12\nabla V_{\mathrm{s}}\cdot \nabla V_{%
\mathrm{t}}^{*}-|V_{\mathrm{s}}|^2V_{\mathrm{s}}V_{\mathrm{t}}^{*}+\alpha
|V_{\mathrm{s}}|^4V_{\mathrm{s}}V_{\mathrm{t}}^{*}-\omega V_{\mathrm{s}}V_{%
\mathrm{t}}^{*}\right) d\mathbf{r}  \nonumber  \\
+\mathrm{c.c.}]+\{1 &\rightleftharpoons &2\},  \label{U}
\end{eqnarray}
where the subscript $D$ pertains to the dimension. Here, $\mathrm{c.c.}$
stands for the complex conjugate expression, the integration is assumed over
the overlapping region in a vicinity of the first soliton, and $%
\{1\rightleftharpoons 2\}$ is the symmetric contribution from the second
soliton. Applying the Gauss theorem to the first term in (\ref{U}), one
transforms, in the 2D case, the expression (\ref{U}) into the form 
\begin{eqnarray}
U_D(R,\psi ) &=&\{[-\int \left( \frac 12\nabla ^2V_{\mathrm{s}}+|V_{\mathrm{s%
}}|^2V_{\mathrm{s}}-\alpha |V_{\mathrm{s}}|^4V_{\mathrm{s}}+\omega V_{%
\mathrm{s}}\right) V_{\mathrm{t}}^{*}d\mathbf{r}  \nonumber \\
+\frac 12\int V_{\mathrm{t}}^{*}\left( \mathbf{n\cdot \nabla }\right) V_{%
\mathrm{s}}dl]+\mathrm{c.c.}\}+\{1 &\rightleftharpoons &2\},  
\label{surface}
\end{eqnarray}
where the surface integral term is taken over a closed contour surrounding
the first soliton, $\mathbf{n}$ being a local vector normal to the contour.
As the contour, one can choose a circumference whose center coincides with
that of the first soliton (Fig. 1). The radius $\rho $ is chosen so that 
\begin{equation}
\kappa ^{-1}\ll \rho \ll R,  \label{rho}
\end{equation}
i.e., it is much larger than the size of the soliton, but much smaller than
the separation between the two solitons. The final objective will be to
obtain an expression that does not depend on the auxiliary radius $\rho $.
To this end, it will be necessary to supplement the condition (\ref{rho}) by
the additional one 
\begin{equation}
\rho ^2/R\ll \kappa ^{-1},  \label{extra}
\end{equation}
which is obviously compatible with (\ref{rho}).

In the 3D case, the difference is that the surface integral in Eq. (\ref
{surface}) is taken over a sphere of the radius $\rho $, so that Fig. 1
shows the central cross section of the 3D situation. The conditions (\ref
{rho}) and (\ref{extra}) pertain equally well to the 3D case.

At this stage of the analysis, the dissipative terms in Eq. (\ref{stationary}%
) are still neglected. Because $V_{\mathrm{s}}$ is an exact single-soliton
solution to Eq. (\ref{stationary}), the first integral term in Eq. (\ref
{surface}) vanishes. The conditions (\ref{rho}) allow one to substitute both 
$V_{\mathrm{t}}$ and $V_{\mathrm{s}}$ in the surface integral term in Eq. (%
\ref{surface}) by the asymptotic expressions (\ref{2D}), which yields, in
the 2D case, 
\begin{eqnarray}
U_2(R,\psi ) &=&-\sqrt{-\frac \omega 2}\left| A_s\right| ^2\sqrt{\rho }%
e^{-\kappa \rho }[\int_0^{2\pi }r^{-1/2}\exp \left( i\psi +is_1\theta
-is_2\eta \right) \exp \left( -\kappa r\right) d\theta  \nonumber  
\\
+\mathrm{c.c.}]+\{1 &\rightleftharpoons &2\}.  \label{12}
\end{eqnarray}
The angles $\eta $ and $\theta $ and the radius 
\begin{equation}
r=\sqrt{(R+\rho \cos \theta )^2+\rho ^2\sin ^2\theta }=R+\rho \cos \theta
+\frac 12\left( \rho ^2/R\right) \sin ^2\theta +...  \label{r}
\end{equation}
are defined in Fig.1, the condition (\ref{rho}) being used to expand the
radical in (\ref{r}). Substituting the expansion into (\ref{12}) and taking
into regard the condition (\ref{extra}), in the first approximation it is
enough to keep the first two terms from (\ref{r}) in $\exp \left( -\kappa
r\right) $, and only the first term in $r^{-1/2}$. Additionally, in the same
approximation one may set $\eta =0$, which leads to 
\begin{eqnarray}
U_2(R,\psi ) &=&-\sqrt{-\frac \omega 2}\left| A_s\right| ^2R^{-1/2}\sqrt{%
\rho }e^{-\kappa \rho }[e^{-\kappa R}\int_0^{2\pi }\exp \left( i\psi
+is_1\theta \right) \exp \left( -\kappa \rho \cos \theta \right) d\theta 
\nonumber  \\
+\mathrm{c.c.}]+\{1 &\rightleftharpoons &2\}.  \label{Bessel}
\end{eqnarray}

The integral in (\ref{Bessel}) can be calculated exactly in terms of the
Bessel functions, but this is not necessary. Indeed, taking into regard, in
line with the previous approximations, that $\kappa \rho \gg 1$, the Laplace
approximation can be applied to the integral, a dominant contribution coming
from a vicinity of the point $\theta =\pi $ (point $A$ in Fig. 1): 
\begin{equation}
\int_0^{2\pi }\exp \left( -\kappa \rho \cos \theta \right) d\theta \approx 
\sqrt{2\pi }(\kappa \rho )^{-1/2}e^{+\kappa \rho }.  \label{Laplace}
\end{equation}
Substituting this into (\ref{Bessel}), one sees that the $\rho $-dependent
multipliers in (\ref{Bessel}) are \emph{exactly} cancelled by $\rho
^{-1/2}e^{+\kappa \rho }$ from (\ref{Laplace}). This cancellation (in the
lowest-order approximation considered here) is a crucial result, as it makes
the effective potential independent of the auxiliary radius $\rho $.

Of course, the dependence on $\rho $ will not disappear if one tries to
calculate higher-order corrections (with respect to $R^{-1}$) to the
effective potential. Actually, this implies that the effective interaction
potential, treating the solitons as particles, can be consistently defined
only in the lowest-order approximation. At the higher orders, it is
necessary to explicitly take into regard deformation of the solitons by the
interactions, which is not an objective of the present work.

In the term $\{1\rightleftharpoons 2\}$ in (\ref{Bessel}), $\psi $ is
replaced, according to its definition, by $-\psi $, and the dominant point
in the surface integral is $\theta =0$. This means that the term $%
\{1\rightleftharpoons 2\}$ is obtained by the change $\psi \rightarrow -\psi 
$, $s_1\pi \rightarrow -s_1\pi $. Finally, in the multiplier $e^{-\kappa R}$
in (\ref{Bessel}), small $q=-\mathrm{Im}\kappa $ should be also taken into
regard (see Eq. (\ref{kappa})), as it gives rise to an important effect,
viz., long-period oscillations in the exponentially decaying tail of the
interaction potential \cite{BS}. Note that the potential does not directly
take into account the model's small dissipative part; however, that part
indirectly affects the potential, inducing the oscillations in the solitons'
tails in (\ref{2D}) through $\mathrm{Im}\kappa $.

With regard to what was said above, the final expression for the potential (%
\ref{Bessel}) is 
\begin{equation}
U_2(R,\psi )=-2\sqrt{2\pi }\left| A_s\right| ^2(-1)^s\cos \psi \;\left( 
\sqrt{-2\omega }/R\right) ^{1/2}\exp \left( -\sqrt{-2\omega }R\right) \cos
(qR),  \label{2Dfinal}
\end{equation}
where $s$ is either $s_1$ or $s_2\equiv \pm s_1$, both giving the same
value. Except for the factors $\left( \sqrt{-2\omega }/R\right) ^{1/2}$ and $%
(-1)^s$, which are specific for the 2D case, the potential (\ref{2Dfinal})
is essentially the same as that obtained in the similar 1D models in \cite
{BS}.

In the 3D case, the consideration is also limited to the interaction of
identical solitons (as it was said above, only the spinless solitons are
considered in the 3D case). The above expression (\ref{surface}) yields the
interaction potential in the 3D case too (recall that, in this case, the
integration in the surface term is over the sphere). As well as in the 2D
case, the first term in (\ref{surface}) vanishes in the approximation that
neglects the direct influence of the dissipation, and the integration over
the sphere is dominated by a contribution from a small vicinity of the point 
$A$ (Fig. 1). Substituting into Eq. (\ref{surface}) the 3D asymptotic
expressions (\ref{3D}) for $V_{\mathrm{s}}$ and $V_{\mathrm{t}}$ and the
expansion (\ref{r}), one arrives, instead of the integral (\ref{Laplace}),
at 
\begin{equation}
2\pi \int_0^\pi \exp \left( -\kappa \rho \cos \theta \right) \sin \theta
\;d\theta =2\pi (\kappa \rho )^{-1}\left( e^{+\kappa \rho }-e^{-\kappa \rho
}\right) \approx 2\pi (\kappa \rho )^{-1}e^{+\kappa \rho }.
\label{3DLaplace}
\end{equation}
With regard to (\ref{3DLaplace}), the final expression for the effective
interaction potential in the 3D case becomes (cf. (\ref{2Dfinal})) 
\begin{equation}
U_3(R,\psi )=-4\pi \left| A_s\right| ^2\cos \psi \;R^{-1}\exp \left( -\sqrt{%
-2\omega }R\right) \cos (qR).  \label{3Dfinal}
\end{equation}
Note that the auxiliary radius $\rho $ is completely canceled out in the
final expressions (\ref{2Dfinal}) and (\ref{3Dfinal}).

The potentials (\ref{2Dfinal}) and (\ref{3Dfinal}) can be as well applied to
the description of the interaction between the 2D and 3D spatiotemporal
solitons (LBs), given by Eqs. (\ref{LB}) and (\ref{2D}), (\ref{3D}). The
differences from the above results are that $q=0$ (recall the dissipation
was completely neglected in the LB models), $\omega $ must be replaced by $-k
$, and the separation $R$ between the solitons is replaced by the \emph{%
spatiotemporal separation} $\Xi $ defined according to Eq. (\ref{LB}): 
\begin{equation}
\Xi =\sqrt{R_{\bot }^2+T^2},  \label{Xi}
\end{equation}
$R_{\bot }$ and $T$ being, respectively, the separation between the solitons
in the transverse direction and the temporal delay between them. 

\subsection{The Model with the Quadratic Nonlinearity}

The interaction potential for the $\chi ^{(2)}$ solitons has its own
peculiarities. For the spatial (stationary) 2D $\chi ^{(2)}$ solitons, the
exponentially decaying potential with two components, generated by the FH
and SH fields, was postulated in \cite{SKB}. The interaction between the $%
\chi ^{(2)}$ LBs is more complicated, because the nonstationary model (\ref
{FH}) and (\ref{SH}) is, effectively, \emph{spatiotemporally anisotropic},
as it was explained in detailed in the previous section, see Eqs. (\ref
{vasymp}), (\ref{wasymp}), and (\ref{locked}). A straightforward
consideration demonstrates that, in both 2D and 3D cases, the SH-generated
interaction potential dominates at $\gamma <\delta $, so that the potential
is given by Eqs. (\ref{2Dfinal}) and (\ref{3Dfinal}), with $\omega $
replaced by $-\gamma $, $q=0$, and $R$ replaced by $\widetilde{\Xi }\equiv 
\sqrt{R_{\bot }^2+\delta ^{-1}T^2}$, cf. Eqs. (\ref{wasymp}) and (\ref{Xi}).
On the contrary to this, at $\gamma >1$, the FH-generated interaction always
dominates, which means that one should use the potentials (\ref{2Dfinal})
and (\ref{3Dfinal}), with $\omega =-1$ and $R$ replaced by $\Xi $ defined as
per Eq. (\ref{Xi}). In the intermediate case $\delta <\gamma <1$ (cf. Eq. (%
\ref{inter2}); recall that the physically relevant case is $\delta <1$), the
interaction potential turns out to be truly anisotropic in the plane $%
(R_{\bot },T)$: the SH-generated interaction dominates inside the sector
(cf. Eq. (\ref{sector2})) 
\begin{equation}
(T/R_{\perp })^2<\frac{1-\gamma }{\gamma -\delta }\delta ,  \label{sector1}
\end{equation}
and the FH-generated interaction dominates outside the sector (\ref{sector1}%
). Accordingly, one should substitute $R$ in the expressions (\ref{2Dfinal})
and (\ref{3Dfinal}) for the interaction potential by $\widetilde{\Xi }$
inside the sector (\ref{sector1}), and by $\Xi $ outside of it.

\section{CONCLUDING REMARKS}

The effective interaction potentials (\ref{2Dfinal}) and (\ref{3Dfinal}) can
give rise to bound states (BS's) of two solitons. In the presence of the
dissipation and gain, it makes sense to consider only BS's of quiescent
solitons, as any motion is suppressed by a friction force. Because the form
of the potentials is essentially the same as in 1D, the situation is not
different from the 1D case, which was recently studied in detail \cite{Seva}%
. There are two types of BS's, with the phase difference between the
solitons $\psi =0$ or $\pi $, and with $\psi =\pi /2$. The BS's of the
former type are saddles, while the BS's of the latter type have imaginary
stability eigenvalues. The fact that the BS's with $\psi =0$ or $\pi $ are
saddles is related to a fundamental property of the interacting solitons:
while an effective mass, $m_R$, of the two-soliton system corresponding to
the radial degree of freedom $R$ is positive, an effective mass $m_\psi $ of
the phase degree of freedom is \emph{negative }\cite{Seva}.

Thus, these two types of the BS's are, respectively, unstable and stable, in
the first approximation. In \cite{Seva}, it has also been demonstrated that
the BS with $\psi =\pi /2$ is subject, in the next approximation, to an
extremely weak instability, which transforms it into a very slowly unwinding
spiral. However, it was also demonstrated that, even if this next-order
instability can be observed, it does not destroy the BS, but, instead, makes
it dynamical, with $R$ and $\psi $ very slowly oscillating in a \emph{limited%
} range. Note that the same mechanism gives rise, in the 1D case, to
(almost) stable chains of the bound solitons; in the 2D and 3D cases, a new
possible pattern is a \emph{lattice} of the bound solitons. There may also
exist ``covalent soliton molecules'', in the form of triangles and
tetrahedrons in the 2D and 3D cases, respectively.

In the absence of the dissipation, BS of mutually orbiting solitons is
possible in the 2D and 3D cases (in the latter case, it is assumed that the
two solitons move in one plane). Orbiting of incoherently interacting 2D
solitons was experimentally observed in a photorefractive medium \cite{Moti}%
. Numerical simulations and analytical arguments presented in \cite{SKB}
demonstrate that the orbiting BS states of the 2D solitons in the $\chi
^{(2)}$ model are unstable. In the present class of the models, the orbiting
BS cannot be stable either. Indeed, for the orbiting state the interaction
potential (\ref{2Dfinal}) or (\ref{3Dfinal}) must be supplemented by the
centrifugal energy $E_{\mathrm{cf}}=\left( M^2/2m_R\right) R^{-2}$, where $M$
is the angular moment of the soliton pair, and $m_R$ is the above-mentioned
effective mass. Thus, the net effective energy of the orbiting state is 
\begin{equation}
E_{\mathrm{eff}}=U_D(R,\psi )+E_{\mathrm{cf}}\equiv C_D\cos \psi
\;R^{-(D-1)/2}\;\exp \left( -\sqrt{-2\omega }R\right) +\left(
M^2/2m_R\right) R^{-2},  \label{effective}
\end{equation}
where, according to Eqs. (\ref{2Dfinal}) and (\ref{3Dfinal}), the constant $%
C_D$ depends on the dimension $D$ and the soliton's spin $s$. It is easy to
check that the effective energy (\ref{effective}) gives rise to a stationary
state with $\sin \psi =0$, $C_D\cos \psi <0$, provided that $M^2$ is small
enough. However, this stationary state always has $\partial ^2E_{\mathrm{eff}
}/\partial R^2<0$, i.e., it is a \emph{maximum} of the effective energy,
consequently, the orbiting BS is unstable against variation of $R$.
Moreover, one can check that the same state always has $\partial ^2E_{%
\mathrm{eff}}/\partial \psi ^2>0$.With regard to the above-mentioned $m_\psi
<0$, this BS is also unstable against variation of $\psi $.

In conclusion, a general method to find the effective potential of
interaction between two-dimensional and three-dimensional solitons was
elaborated, including the case of the two-dimensional vortex (spinning)
solitons. The method is based on calculation of the overlapping term in the
full Hamiltonian of the system. The main technical point that makes the
calculation possible is that the bulk integral reduces to a surface one,
and, in the lowest-order approximation, the final expression does not
contain the auxiliary radius of the overlapping region. The result applies
to spatial solitons and ``light bullets'' (spatiotemporal solitons) in
nonlinear optics (in the model with the quadratic nonlinearity, the
interaction between the ``bullets'' may be spatiotemporally anisotropic).
The interaction potential predicts that an orbiting bound state of two
solitons exists, but is always unstable. In the presence of weak dissipation
and gain, the effective potential can also be derived, giving rise to bound
states of the solitons (both unstable and almost stable) similar to those
recently studied in the one-dimensional model.

\newpage

\section{FIGURE\ CAPTION}

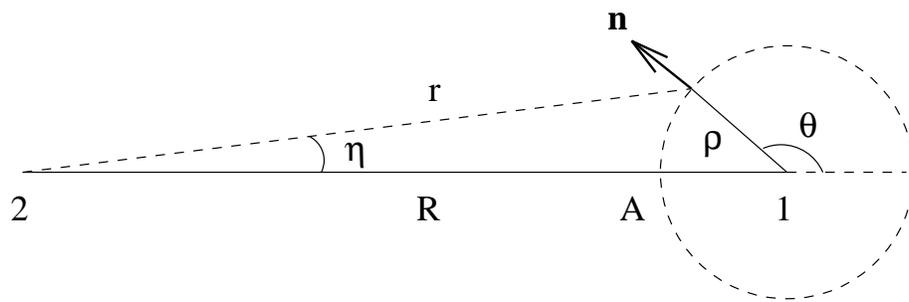
Fig. 1. The two-soliton configuration in the two- and three-dimensional
cases (in the latter case, the figure shows the central cross section of the
3D configuration). The points $1$ and $2$ are centers of the solitons.


\newpage

\begin{thebibliography}{99}
\bibitem{Manolo}  M. Quiroga-Teixeiro and H. Michinel, J. Opt. Soc. Am. B 
\textbf{14}, 2004 (1997).

\bibitem{SKB}  V.V. Steblina, Yu.S. Kivshar, and A.V. Buryak, Opt. Lett. 
\textbf{23}, 156 (1998).

\bibitem{Moti}  M. Shih, M. Segev, and G. Salamo, Phys. Rev. Lett. \textbf{78%
}, 2551 (1997).

\bibitem{Moti2}  H. Meng, G. Salamo, and M. Segev, Opt. Lett. \textbf{23},
897 (1998).

\bibitem{Krolik}  W. Kr\'{o}likowski, M. Saffman, B. Luther-Davies, and C.
Denz, Phys. Rev. Lett. \textbf{80}, 3240 (1998).

\bibitem{KR}  A.A. Kanashov and A.M. Rubenchik, Physica D \textbf{4,} 122
(1981).

\bibitem{Yaron}  Y. Silberberg, Opt. Lett. \textbf{15}, 1281 (1990).

\bibitem{Aussie}  B.A. Malomed, P. Drummond, H. He, D. Anderson, A.
Berntson, and M. Lisak, Phys. Rev. E \textbf{56, }4725 (1997).

\bibitem{Torner}  D. Mihalache, D. Mazilu, B.A. Malomed, and L. Torner, Opt.
Comm. \textbf{152}, 365 (1998).

\bibitem{BS}  B.A.~Malomed, Phys.~Rev.~A \textbf{44}, 6954 (1991);
B.A.~Malomed, Phys.~Rev.~E \textbf{47}, 2874 (1993); B.A. Malomed and A.A.
Nepomnyashchy, Europhys. Lett. \textbf{27}, 649 (1994).

\bibitem{PTS}  B.L. Lawrence, M. Cha, J.U. Kang, W. Torruellas, G. Stegeman,
G. Baker, J. Meth, and S. Etemad, Electr. Lett. \textbf{30}, 889 (1994).

\bibitem{PS}  V.I. Petviashvili and A.M. Sergeev, Dokl. AN\ SSSR \textbf{276}%
, 1380 (1984) [Sov. Phys. Doklady \textbf{29}, 493 (1984)].

\bibitem{CH}  P. Marcq, H. Chat\'{e}, and R. Conte, Physica D \textbf{73},
305 (1994).

\bibitem{1D}  B.A.~Malomed, Physica D \textbf{23,} 155 (1987).

\bibitem{3}  W.~van~Saarloos and P.C.~Hohenberg, Phys. Rev. Lett. \textbf{64,%
} 749 (1990); V.~Hakim, P.~Jakobsen, and Y.~Pomeau, Europhys. Lett. \textbf{%
11,} 19 (1990); B.A. Malomed and A.A. Nepomnyashchy, Phys. Rev. A \textbf{42}%
, 6009 (1990).

\bibitem{Aranson}  I.S. Aranson, K.A. Gorshkov, A.S. Lomov, and M.I.
Rabinovich, Physica D \textbf{43}, 435 (1990).

\bibitem{Ll}  D.V. Petrov and L. Torner, Opt. Quant. Electr.\textbf{\ 29},
1037 (1997).

\bibitem{GO}  K.A. Gorshkov, L.A. Ostrovsky, and E.N. Pelinovsky, Proc. IEEE 
\textbf{82}, 1511 (1974).

\bibitem{Seva}  V.V. Afanasjev, P.L. Chu, and B.A. Malomed, Phys. Rev. E 
\textbf{57}, 1088 (1998).
\end{thebibliography}
\end{document}